\documentclass{aastex631}
\usepackage[utf8]{inputenc}
\usepackage{CJK}
\usepackage{amsmath}

\usepackage{booktabs}
\usepackage{tablefootnote}
\usepackage{multirow}

\newcommand{\mum}{\ensuremath{\,{\rm \mu m}}}

\newcommand{\K}{\ensuremath{\,{\rm K}}}

\newcommand{\s}{\ensuremath{\,{\rm s}}}
\newcommand{\g}{\ensuremath{\,{\rm g}}}

\newcommand{\citealias}[1]{\cite{Weingartner2001} (WD)!)}

\shorttitle{IR Extinction in Dense Clouds}
\shortauthors{Li et al.}

\begin{document}
\begin{CJK*}{UTF8}{gbsn}
		
\title{The Flattest Infrared Extinction Curve in Four Isolated Dense Molecular Cloud Cores}
		
\author[0000-0001-9328-4302]{Jun Li (李军)}
\affiliation{Center for Astrophysics, Guangzhou University, Guangzhou 510006, People's Republic of China}
		
\author[0000-0003-2472-4903]{Bingqiu Chen (陈丙秋)}
\affiliation{South-Western Institute for Astronomy Research, Yunnan University, Kunming, Yunnan 650091, Peopleʼs Republic of China}
		
\author[0000-0003-3168-2617]{Biwei Jiang (姜碧沩)}
\affiliation{Institute for Frontiers in Astronomy and Astrophysics, Beijing Normal University, Beijing 102206,  People's Republic of China}
\affiliation{School of Physics and Astronomy, Beijing Normal University, Beijing 100875, People's Republic of China}
		
\author[0000-0003-2645-6869]{He Zhao (赵赫)}
\affiliation{Purple Mountain Observatory and Key Laboratory of Radio Astronomy, Chinese Academy of Sciences, 10 Yuanhua Road, Nanjing 210033, People's Republic of China}
		
\author[0009-0007-2880-849X]{Botao Jiang (江博韬)}
\affiliation{Center for Astrophysics, Guangzhou University, Guangzhou 510006, People's Republic of China}
		
\author[0000-0002-5435-925X]{Xi Chen (陈曦)}
\affiliation{Center for Astrophysics, Guangzhou University, Guangzhou 510006, People's Republic of China}
		
\correspondingauthor{Jun Li}
\email{lijun@gzhu.edu.cn}

\received{25 September 2024}
\revised{30 October 2024}
\accepted{31 October 2024}

\begin{abstract}
The extinction curve of interstellar dust in the dense molecular cloud cores is crucial for understanding dust properties, particularly size distribution and composition. We investigate the infrared extinction law in four nearby isolated molecular cloud cores, L429, L483, L673, and L1165, across the 1.2 - 8.0 $\mu$m wavelength range, using deep near-infrared (NIR) and mid-infrared (MIR) photometric data from UKIDSS and Spitzer Space Telescope. These observations probe an unprecedented extinction depth, reaching $A_V\sim$ 40-60 mag in these dense cloud cores. We derive color-excess ratios $E(K-\lambda)/E(H-K)$ by fitting color-color diagrams of $(K-\lambda)$ versus $(H-K)$, which are subsequently used to calculate the extinction law $A_\lambda/A_K$. Our analysis reveals remarkably similar and exceptionally flat infrared extinction curves for all four cloud cores, exhibiting the most pronounced flattening reported in the literature to date. This flatness is consistent with the presence of large dust grains, suggesting significant grain growth in dense environments. Intriguingly, our findings align closely with the Astrodust model for a diffuse interstellar environment proposed by Hensley \& Draine. This agreement between dense core observations and a diffuse medium model highlights the complexity of dust evolution and the need for further investigation into the processes governing dust properties in different interstellar environments.
\end{abstract}
		
		
\section{Introduction} \label{sec:intro}
The central regions of dark molecular clouds in the Milky Way typically exhibit extremely low temperatures ($\sim10\,\rm K$) and exceptionally high densities ($10^4\text{-}10^6\,\rm cm^{-3}$), and are often the birthplaces of stars and their planetary systems \citep{Bergin2007}. Accurately measuring and understanding these initial physical conditions of dark cloud cores is crucial for unraveling the fundamental nature of the star formation process \citep{Andre2014,Offner2014}. However, molecular hydrogen, the primary component of star formation, is difficult to observe directly under such low-temperature and high-density conditions. Consequently, the study of dust becomes essential for probing the physical conditions and structure of these dark cloud cores \citep{Lombardi2014,Sadavoy2018}. 
		
The extinction law is a crucial tool for studying dust properties, as it can constrain key information such as dust size distribution and chemical composition \citep{Draine2003}. In the optical and ultraviolet wavelengths, the extinction law has been successfully applied to investigate dust characteristics in the diffuse interstellar medium (ISM), although the shape parameter of the extinction curve, $R_V$ (the ratio of extinction $A_V$ to color excess $E(B-V)$), varies significantly along different lines of sight \citep{Cardelli1989}. In the near-infrared (NIR, 1-3\,$\mu\rm m$), the extinction law is considered to be a universal and invariant power-law form ($A_\lambda\propto\lambda^{-\alpha}$), with the spectral index $\alpha$ varying between different environments, approximately ranging from 1.6 to 2.4 (see \citealt{Matsunaga2018} for a review). The NIR extinction law is also reported to vary with wavelength \citep{Hosek2018,Nogueras2019}. For the mid-infrared (MIR, 3-8\,$\mu\rm m$) extinction law, a clear description or theoretical explanation has not yet been obtained \citep{Gao2009}. However, in high optical depth regions, it becomes extremely challenging, as optical and ultraviolet observations cannot penetrate these areas. Therefore, the study of the infrared extinction law becomes particularly important.

Furthermore, in dense environments, dust grains continuously grow through the processes of coagulation and the formation of ice mantle, facilitating their adhesion and subsequent growth \citep{Ossenkopf1994,Ormel2011}. This process not only leads to an increase in the far-infrared emission coefficient of dust \citep{Stepnik2003,Li2024a} but also significantly alters the size distribution of dust grains, causing the observed extinction law to deviate from the typical form found in diffuse ISM \citep{Draine2003}. In other words, the dust extinction law tends to be higher in dense interstellar environments. Interestingly, however, numerous observations indicate that the MIR extinction law, spanning the 3-8 $\mu$m wavelength range, exhibits a relatively flat characteristic, regardless of whether in diffuse or dense environments \citep{Indebetouw2005,Flaherty2007,Li2024b}. These observational results are quite close to the classical $R_V=5.5$ dust extinction model curve and sometimes even show a flatter shape.

In recent years, studies of the MIR extinction law have primarily focused on the Galactic center or near Galatic disk regions \citep{McClure2009,Gao2009,Fritz2011}. However, these regions are typically influenced by a complex interplay of various physical conditions along the line of sight, such as non-uniform interstellar radiation fields, mixing of multi-phase ISM, and stellar feedback processes. The combined effects of these factors make it extremely challenging to accurately analyze the relationship between dust properties and their environment. \cite{Gao2009} and \cite{Zasowski2009} utilized Spitzer data and discovered significant spatial variations in the MIR extinction law on a galactic scale. In contrast, isolated molecular clouds are far from strong radiation fields, and their internal physical conditions are relatively simple, with minimal external interference \citep{Clemens1991,Kandori2005}. These characteristics make isolated molecular clouds ideal laboratories for studying the dust properties.
		
In this work, we study the infrared extinction law in four isolated molecular cloud cores: L429, L483, L673, and L1165 by using NIR photometric observations from United Kingdom Infrared Telescope (UKIRT) Infrared Deep Sky Survey (UKIDSS) and MIR observations from Spitzer. These four clouds are relatively nearby and isolated, ensuring that any observed extinction characteristics can be attributed solely to the dust associated with each individual cloud. Section \ref{sec:data} details the data used in this work. We present and discuss the observed extinction laws in Section \ref{sec:results}, followed by a summary in Section \ref{sec:summary}.
		
\section{Data} \label{sec:data}
We select four sources from the nearby isolated molecular cloud cores of \cite{Sadavoy2018}. These sources have both NIR $JHK$ band observations from UKIDSS and MIR observations from Spitzer. They represent different evolution stages: L429 and L673 are collapsing starless cores \citep{Stutz2009,Sepulveda2011}, L483 has Class 0 protostars \citep{Fuller1995}, while L1165 contains Class I protostars \citep{Tapia1997}. Table \ref{tab:sample} lists the basic parameters of the four cloud cores studied in this work, including their coordinates, distances, sizes, total numbers of stars in the UKIRT field of view, and maximum extinction $A_J$ determined by \cite{Juvela2016} using 2MASS survey. These cloud cores are all located in the solar vicinity, with distances ranging from 200 to 300 pc and spatial scales between 0.5 and 1.2 pc. These cloud cores are relatively isolated and less influenced by their surrounding environments, making them ideal targets for studying the properties of dust in dense interstellar environments.
		
\subsection{UKIDSS Data}
We use NIR data from the UKIDSS Galactic Plane Survey (GPS) Data Release (DR) 11 \citep{Lucas2008}. The UKIDSS GPS employs the UKIRT to conduct large-scale surveys of the Galactic plane in three bands: $J$ (1.25\mum), $H$ (1.63\mum), and $K$ (2.19\mum). These data can be obtained through the Wide Field Camera (WFCAM) Science Archive\footnote{http://surveys.roe.ac.uk/wsa/index.html}. To ensure a data catalog with high reliability, we follow the data quality control methods recommended by \cite{Lucas2008}, which include removing blended objects, sources with saturated pixels, and mismatched sources. However, we apply a different photometric error cut of 0.2 mag in each photometric filter. Notably, the UKIDSS photometric depth surpasses that of 2MASS by more than 3 mag in each band.
		
\subsection{Spitzer Data}
We use MIR data from the Spitzer ``From Molecular Cores to Planet-Forming Disks" project (c2d; \citealt{c2d}). The photometry are obtained by point spread function (PSF) fitting. Detailed descriptions of the c2d project and data processing methods can be found in \cite{Evans2003} and through the Spitzer Website\footnote{https://irsa.ipac.caltech.edu/data/SPITZER/C2D/doc/c2d\_del\_document.pdf}. Spitzer c2d data is available through the archive on the Spitzer Science Center\footnote{https://irsa.ipac.caltech.edu/data/SPITZER/C2D/}. We use the photometric data from the four bands of Spitzer Infrared Array Camera (IRAC), namely [3.6], [4.5], [5.8], and [8.0] $\mu\rm m$. To ensure the reliability of the photometric data, we apply strict quality control to the c2d data. We require that the sources have a Signal-to-Noise Ratio (SNR) greater than 3 to guarantee the photometric accuracy. Furthermore, we only select sources that are flagged as ``star" in the ``OType" column of the c2d catalog to minimize contamination from background galaxies and image artifacts. We cross-match the UKIDSS and Spitzer data using a 1$''$ radius, resulting in a sample of approximately several hundred to a thousand stars, as listed in Table \ref{tab:sample}.
		
\begin{table*}
\centering
\caption{Properties of the four dense cloud cores.}
\label{tab:sample}
\begin{tabular}{c c c c c c c}
\hline \hline
Cores &  R.A., Dec. (J2000) & Distance (pc) & Size (pc)& No. of stars & max $A_J$ (mag)$^a$ & Evolutionary Stage  \\
\hline
L429 & 18:17:05.5, $-$08:14:41   & 225 $\pm$ 55 (1)  & 0.6 & 920 & 3.4  & Collapsing, starless (2)\\
L483   & 18:17:29.9, $-$04:39:41  & 225 $\pm$ 55 (1)  & 0.5& 413 & 2.4  & Class 0 star-forming (3)  \\
L673  & 19:20:25.3, $+$11:22:14    & 260 $\pm$ 50 (4) & 0.9&1041 & 3.3  & Collapsing, starless (5) \\
L1165 & 22:06:50.6, $+$59:02:43 & 300 $\pm$ 50 (6) & 1.2  & 436& 1.8  & Class I star-forming (7) \\
\hline
\end{tabular}
\tablecomments{(a) Maximum $J$-band extinction ($A_J$) within the field of view of each dense core, derived from the 2MASS all-sky extinction map \citep{Juvela2016}. \\
References: (1) \cite{Straizys2003}; (2)\cite{Stutz2009}; (3) \cite{Fuller1995}; (4) \cite{Das2015}; (5) \cite{Sepulveda2011}; (6) \cite{Gyulbudagya1985}. (7) \cite{Tapia1997}}
\end{table*}

\section{Results and Discussion} \label{sec:results}
		
\subsection{The NIR Color-Excess Ratio $E(J-H)/E(H-K)$}\label{sec:nir_ratios}
		
\begin{figure*}[ht!]
\centering
\includegraphics[scale=0.5]{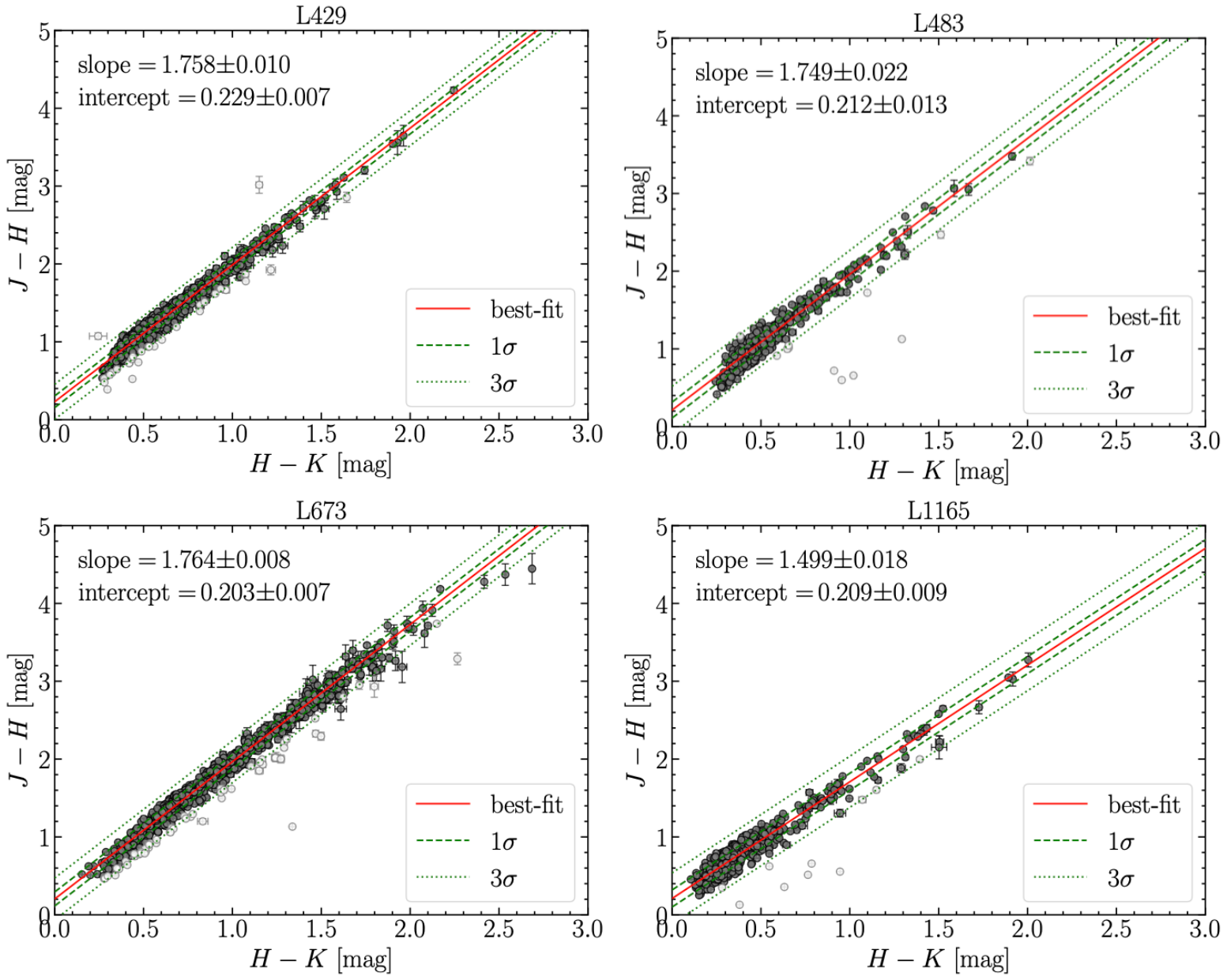}
\caption{UKIDSS color-color diagrams $(J-H)$ versus $(H-K)$ for the four dense cloud cores L429, L483, L673, and L1165. The red solid lines represent the best-fit linear relations obtained using the \texttt{LTS\_LINEFIT} method, while the green dashed and dotted lines indicate the 1$\sigma$ and 3$\sigma$ confidence intervals of the fits, respectively. The grey markers denote outliers identified by the \texttt{LTS\_LINEFIT} algorithm. Best-fit parameters, including the slope and intercept, are displayed in the upper-left corner of each panel. \label{fig:nir_ratios}}
\end{figure*}

We first construct color-color diagrams of $(J-H)$ versus $(H-K)$ for the four dense clouds using the high-quality photometric catalog, as shown in Figure \ref{fig:nir_ratios}. These diagrams serve as powerful tools for analyzing the extinction properties in the NIR regime. The stars observed in the NIR color-color diagrams are predominantly giants located behind the cloud core. Given that the intrinsic color dispersion of giants in the NIR is typically small (less than 0.1 mag) \citep{Koornneef1983,Lombardi2011} and well-constrained photometric uncertainties, the overall linear distributions in Figure \ref{fig:nir_ratios} are remarkably tight. This linear distribution is primarily attributed to dust reddening, allowing for the direct derivation of the NIR color-excess ratio $E(J-H)/E(H-K)$ through linear fitting the color-color diagrams.

To determine the color-excess ratio $E(J-H)/E(H-K)$ through a robust fit to the color-color distribution of $(J-H)$ versus $(H-K)$, we employed the Python package LTS\_LINEFIT\footnote{https://pypi.org/project/ltsfit} \citep{Cappellari2014} for linear regression. LTS\_LINEFIT accounts for the errors in both color axes and the intrinsic scatter of the distribution, while automatically identifying outliers. Figure \ref{fig:nir_ratios} presents the fitting results for the four cloud cores, with only a few stars (depicted as grey markers) being excluded. These removed stars lie outside the best-fitting $3\sigma$ range and are potentially young stellar objects (YSOs) or evolved stars with circumstellar dust. The derived NIR color-excess ratios $E(J-H)/E(H-K)$ are $1.758\pm0.010$, 1.749$\pm$0.022, 1.764$\pm$0.008, and 1.499$\pm$0.018 for L429, L483, L673, and L1165, respectively. Notably, three of the four dense clouds exhibit comparable values of approximately 1.75, while L1165 presents a significantly lower value of about 1.5. This discrepancy may indicate a real variation in NIR extinction across different environments. The markedly lower $E(J-H)/E(H-K)$ value for L1165 could potentially be attributed to the presence of young stars and outflows within L1165 \citep{Parker1991,Sepulveda2011}, which may influence dust properties.

Numerous studies have measured the NIR color-excess ratio $E(J-H)/E(H-K)$ in various environments of the ISM. One of the earliest and most widely referenced studies on NIR extinction was conducted by \cite{Rieke1985}. They derived a color-excess ratio of $E(J-H)/E(H-K)\approx$1.70 in the Galactic center. Subsequently, \cite{Indebetouw2005} obtained a very similar result of approximately 1.73 using 2MASS data. Focusing on dark clouds, several studies have found NIR color-excess ratios $E(J-H)/E(H-K)$ ranging between 1.6 and 2.0 \citep[e.g.,][]{Kenyon1998,Racca2002,Naoi2006,Naoi2007,Chen2013}, suggesting spatial variations dependent on the line of sight and local physical conditions of the dust. Our results align closely with these previous findings, showing no significant deviations.

In Appendix \ref{sec:app_nir}, we calculate the NIR extinction parameters including the power-law indices $\alpha$, $A_J/A_K$, and $A_H/A_K$ by assuming a power-law form of NIR extinction, with results presented in Table \ref{tab:nir_cer}. Our analysis reveals that the derived values of $\alpha$, $A_J/A_K$, and $A_H/A_K$ for L429, L483, and L673 are systematically higher than those from \cite{Indebetouw2005} and the WD01 $R_V=5.5$ model, despite similar observed values of $E(J-H)/E(H-K)$. For instance, an $E(J-H)/E(H-K)$ value of 1.75 yields an $\alpha$ of 2.0, and the calculated $A_J/A_K$ and $A_H/A_K$ ratios of 2.9 and 1.7, respectively, which are apparently larger than the values reported by \cite{Indebetouw2005} of 2.50 and 1.55. Furthermore, for the extinction law with  $R_V=5.5$ from the models of \citet[][here after WD01]{Weingartner2001}, the $E(J-H)/E(H-K)$ ratio is 1.77, with $A_J/A_K$ and $A_H/A_K$ values of 2.44 and 1.52, respectively, closely aligning with \cite{Indebetouw2005}. The discrepancies observed in converting from color-excess ratios to power-law indices $\alpha$ and extinction ratios, when compared to WD01 $R_V=5.5$ and \cite{Indebetouw2005}, suggest potential inaccuracies in this conversion process. This conversion appears to be highly sensitive to effective wavelengths. Alternatively, the power-law assumption may not be strictly valid, as exemplified by the WD01 extinction model, which does not exhibit a strictly power-law form in the $JHK$ bands.
		
\subsection{The MIR Reddening Law }
		
\subsubsection{Determination of Color-excess Ratios $E(K-\lambda)/E(H-K)$ }\label{sec:EHK}
		
To determine the color-excess ratios in the MIR, we use $(H-K)$ as the reference color and present color-color diagrams of $(K-\lambda)$ versus $(H-K)$ in Figure \ref{fig:mir_ratios_L673}, and Figures \ref{fig:mir_ratios_L429}-\ref{fig:mir_ratios_L1165} in the Appendix \ref{sec:app_mir_cer}, where $\lambda$ corresponds to the Spitzer/IRAC bands: [3.6], [4.5], [5.8], and [8.0]. Although some previous studies have employed $(J-K)$ as the reference color due to its higher sensitivity to extinction, we adopt for $(H-K)$ as it allows for greater extinction depth in dense molecular clouds. For instance, in the case of L673, comparing Figure \ref{fig:nir_ratios} with Figure \ref{fig:mir_ratios_L673} reveals that the detection depth using the $J$ band reaches only up to $(H-K)\sim$ 2.5 mag, whereas using $(H-K)$ as the reference color extends the depth to 4 mag, which corresponds to an optical extinction depth of exceeding 60 mag.

As discussed in Section \ref{sec:nir_ratios}, the slope of the linear fit to the color distribution represents the color-excess ratio $\beta_\lambda = E(K-\lambda)/E(H-K)$. The stellar color distributions for all four IRAC bands exhibit a clear linear relation between $(K-\lambda)$ and $(H-K)$, though the scatter increases slightly as wavelength increases from 3.6\mum\ to 8.0\mum. We employ the \texttt{LTS\_LINEFIT} package to perform linear fits and derive the color-excess ratios, with results presented in Table \ref{tab:results}. Figure \ref{fig:cer_curve} illustrates the observed color-excess ratios $E(K-\lambda)/E(H-K)$ as a function of wavelength $\lambda$ for the four cloud cores: L429, L483, L673, and L1165. These ratios follow a smooth trend, with only minor variations among the four clouds, particularly in the IRAC bands. Notably, L1165 exhibits slightly larger MIR color-excess ratios, resulting in a marginally steeper curve.

The observed color-excess curves are flatter than both the classical WD01 dust extinction models for $R_V=3.1$ and $R_V=5.5$, which represent extinction in the diffuse and dense ISM, respectively. Moreover, these curves are even flatter than the reddening curve measured by \cite{Hensley2020} for Cyg OB2-12, which also traces the diffuse ISM. Interestingly, our observed color-excess ratios align most closely with the extinction curve predicted by the Astrodust model proposed by \cite{Hensley2023}, despite this model being developed primarily for diffuse environments. The Astrodust model we compare to represents a unified description of interstellar dust consisting of a single composite material (``astrodust"), complemented by smaller polycyclic aromatic hydrocarbons (PAHs). While developed for the diffuse ISM, this model employs typical sub-micron grain size distributions with modified dielectric functions, successfully reproducing observed extinction, emission, and polarization characteristics while satisfying elemental abundance constraints. The remarkable agreement between our dense core observations and this diffuse ISM model suggests that the modified dielectric functions might capture some essential aspects of grain properties that remain relevant even in dense environments, though the underlying physical interpretation requires further investigation.
		
\subsubsection{Determination of $A_\lambda/A_K$}
		
Using the color-excess ratios $\beta_\lambda=E(K-\lambda)/E(H-K)$ derived in Section \ref{sec:EHK}, we can convert these into the extinction law $A_\lambda/A_K$ using the following equation:
\begin{equation}\label{equ:AK}
			\frac{A_\lambda}{A_K}=1-(\frac{A_H}{A_K}-1)\beta_\lambda
\end{equation}
This conversion relies on the NIR extinction ratio $A_H/A_K$. Based on the discussion of NIR extinction law in Section \ref{sec:nir_ratios}, our derived NIR color-excess ratios align most closely with those reported by \cite{Indebetouw2005} and WD01 $R_V=5.5$ model \citep{Weingartner2001}. Consequently, we adopt $A_H/A_K=1.55$ for Equation \ref{equ:AK}. Table \ref{tab:results} presents the derived $A_\lambda/A_K$, while Figure \ref{fig:ext_curve} illustrates the extinction curves $A_\lambda/A_K$ for the four cloud cores, alongside several extinction laws from the literature for comparison. 
		
As evident in Figure \ref{fig:ext_curve}, the observed extinction curves of the four cloud cores are among the flattest reported to date. This pronounced flatness in the MIR region potentially suggests the presence of larger dust grains in these dense environments. This finding aligns with the concept of grain growth through processes such as coagulation and accretion in molecular cloud cores. The comparison with previous studies further accentuates the distinctive properties of dust in the dense regions we analyzed, underscoring the unique nature of these environments.

\begin{table*}
\centering
\caption{Color-excess ratios $E(K-\lambda)/E(H-K)$ and resulting extinction law $A_\lambda/A_K$.}
\label{tab:results}
\begin{tabular}{c c c c c c c c}
\hline \hline
Cores &  $J$ & $H$ &$K$ & $[3.6]$ & $[4.5]$ & [5.8] & [8.0] \\
& 1.250\mum$^a$ & 1.632\mum & 2.193\mum & 3.521\mum & 4.440\mum & 5.670\mum &7.676\mum \\
\hline
\multicolumn{8}{ c }{$E(K-\lambda)/E(H-K)$} \\
\hline
L429 &$-$2.760 $\pm$ 0.010 & 1.0 & 0.0 & 0.423 $\pm$ 0.005& 0.590 $\pm$ 0.006 & 0.699 $\pm$ 0.006 & 0.706 $\pm$ 0.007 \\
L483  & $-$2.752 $\pm$ 0.022 & 1.0 & 0.0 & 0.443 $\pm$ 0.008 & 0.596 $\pm$ 0.010 & 0.715 $\pm$ 0.012 & 0.690 $\pm$ 0.013  \\
L673 &$-$2.756 $\pm$ 0.008 & 1.0 & 0.0 & 0.470 $\pm$ 0.005 & 0.608 $\pm$ 0.005 & 0.719 $\pm$ 0.005 & 0.724 $\pm$ 0.006 \\
L1165 & $-$2.502 $\pm$ 0.019 & 1.0 & 0.0 & 0.516 $\pm$ 0.008 & 0.667 $\pm$ 0.081 & 0.800 $\pm$ 0.011 & 0.794 $\pm$ 0.012 \\
Average & $-$2.693 $\pm$ 0.110& 1.0 & 0.0 & 0.463 $\pm$ 0.035 & 0.615 $\pm$ 0.030 & 0.733 $\pm$ 0.039 & 0.728 $\pm$ 0.040 \\
\hline
\multicolumn{8}{ c }{$A_\lambda/A_K$} \\
\hline
L429 & 2.518 $\pm$ 0.009 & 1.55 & 1.0 & 0.767 $\pm$ 0.009 & 0.676 $\pm$ 0.006 & 0.616 $\pm$ 0.006 & 0.612 $\pm$ 0.006 \\
L483 & 2.514 $\pm$ 0.020 & 1.55 & 1.0 & 0.757 $\pm$ 0.014 & 0.672 $\pm$ 0.011 & 0.607 $\pm$ 0.010 & 0.621 $\pm$ 0.012 \\
L673 & 2.526 $\pm$ 0.007 & 1.55 & 1.0 & 0.741 $\pm$ 0.008 & 0.665 $\pm$ 0.005 & 0.605 $\pm$ 0.004 & 0.602 $\pm$ 0.005 \\
L1165 & 2.376 $\pm$ 0.019 & 1.55 & 1.0 & 0.716 $\pm$ 0.012 & 0.633 $\pm$ 0.008 & 0.560 $\pm$ 0.008 & 0.563 $\pm$ 0.009 \\
Average & 2.481 $\pm$ 0.061 & 1.55 & 1.0 & 0.745 $\pm$ 0.019 & 0.662 $\pm$ 0.017 & 0.597 $\pm$ 0.022 & 0.599 $\pm$ 0.022 \\
\hline
\end{tabular}
\tablecomments{$a$ The effective wavelengths are derived by convolving the filter transmissions of UKIDSS and IRAC with spectrum of a K0III star from \cite{Castelli2003}.}
\end{table*}

\subsection{Comparison with Previous Studies}
		
Numerous measurements based on Spitzer observations have consistently shown that the MIR extinction law appears to be flat in a variety of environments, including both diffuse and dense ISM regions (see \citealt{Matsunaga2018} for a review). Many of these studies have found that the observed extinction laws in dense ISM regions are either comparable to, or even flatter than, the WD01 $R_V$=5.5 model extinction law. For example, \cite{Indebetouw2005} reported a flat extinction law in both diffuse and dense regions of the ISM, while \cite{Flaherty2007} and \cite{McClure2009} found similar results in star-forming regions. \cite{Xue2016} used spectroscopic data to provide precise measurements of the average MIR extinction law in the Milky Way, further supporting the flat nature of MIR extinction in various environments. However, steep MIR extinction laws have also been observed. For example, a study by \cite{Sanders2022} found a steep extinction relationship (low $A_\lambda/A_K$) for inner Galactic regions with typically low $A_V$. Consequently, the nature of MIR extinction laws remains a subject of considerable debate.
		
\begin{figure*}[ht!]
\centering
\includegraphics[scale=0.5]{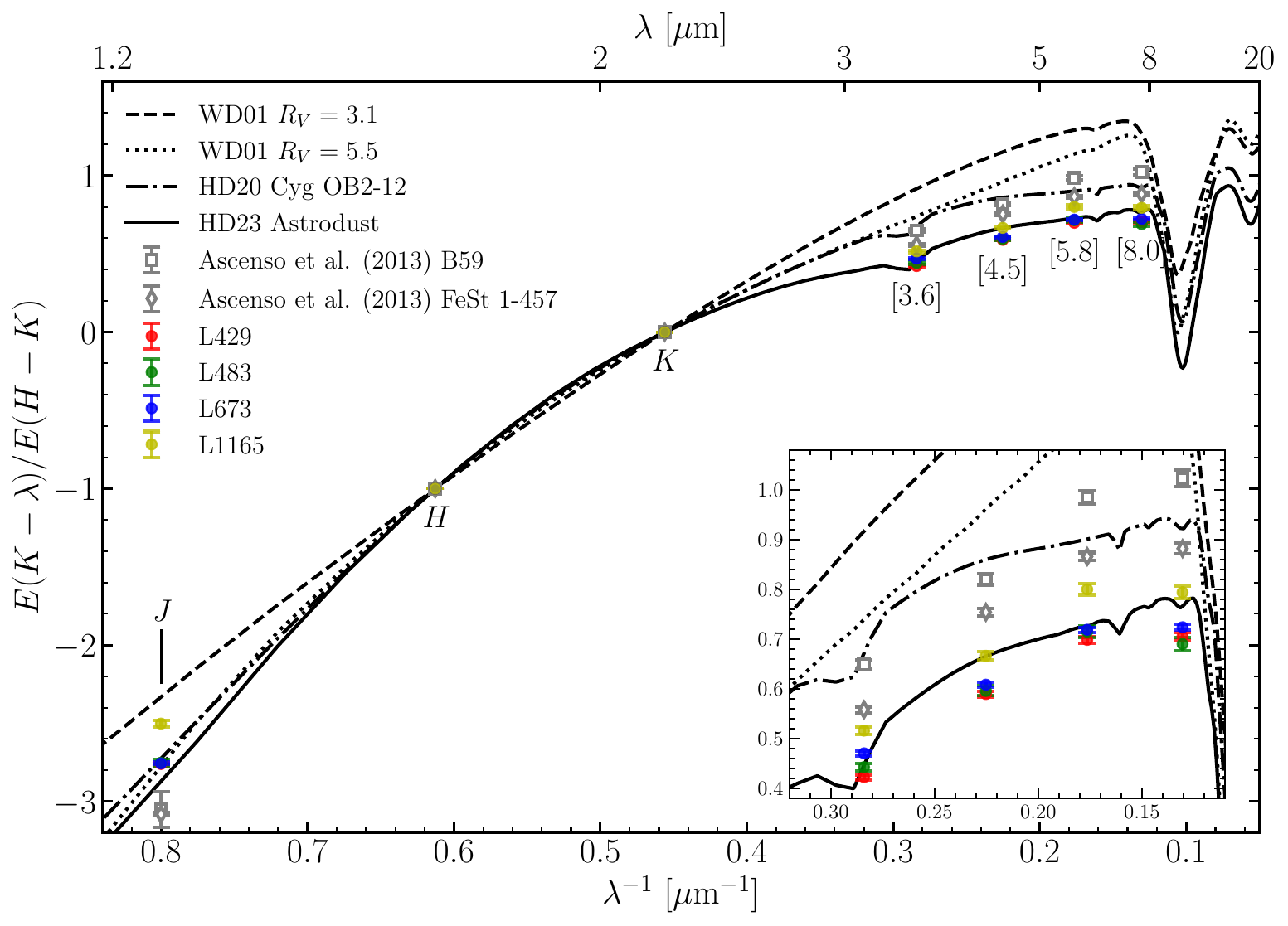}
\caption{Color-excess ratios $\beta_\lambda=E(K-\lambda)/E(H-K)$ as a function of  wavelength for the four dense cloud cores L429, L483, L673, and L1165 are shown with various colored markers. The extinction laws predicted by \citet[][WD01]{Weingartner2001} for $R_V$=3.1 and $R_V$=5.5 are represented by dashed and dotted lines, respectively, while the Astrodust model by \citet[][HD23]{Hensley2023} is depicted with a solid line. For comparison, the observed extinction law for the diffuse ISM toward Cyg OB2-12 is shown with a dot-dashed line from \citet[][HD20]{Hensley2020}. The measurements from \cite{Ascenso2013} for the dense cores B59 and FeSt1-457 are shown in grey markers. The inset is a zoom-in view of the MIR regime. \label{fig:cer_curve}}
\end{figure*}

\begin{figure*}[ht!]
\centering
\includegraphics[scale=0.55]{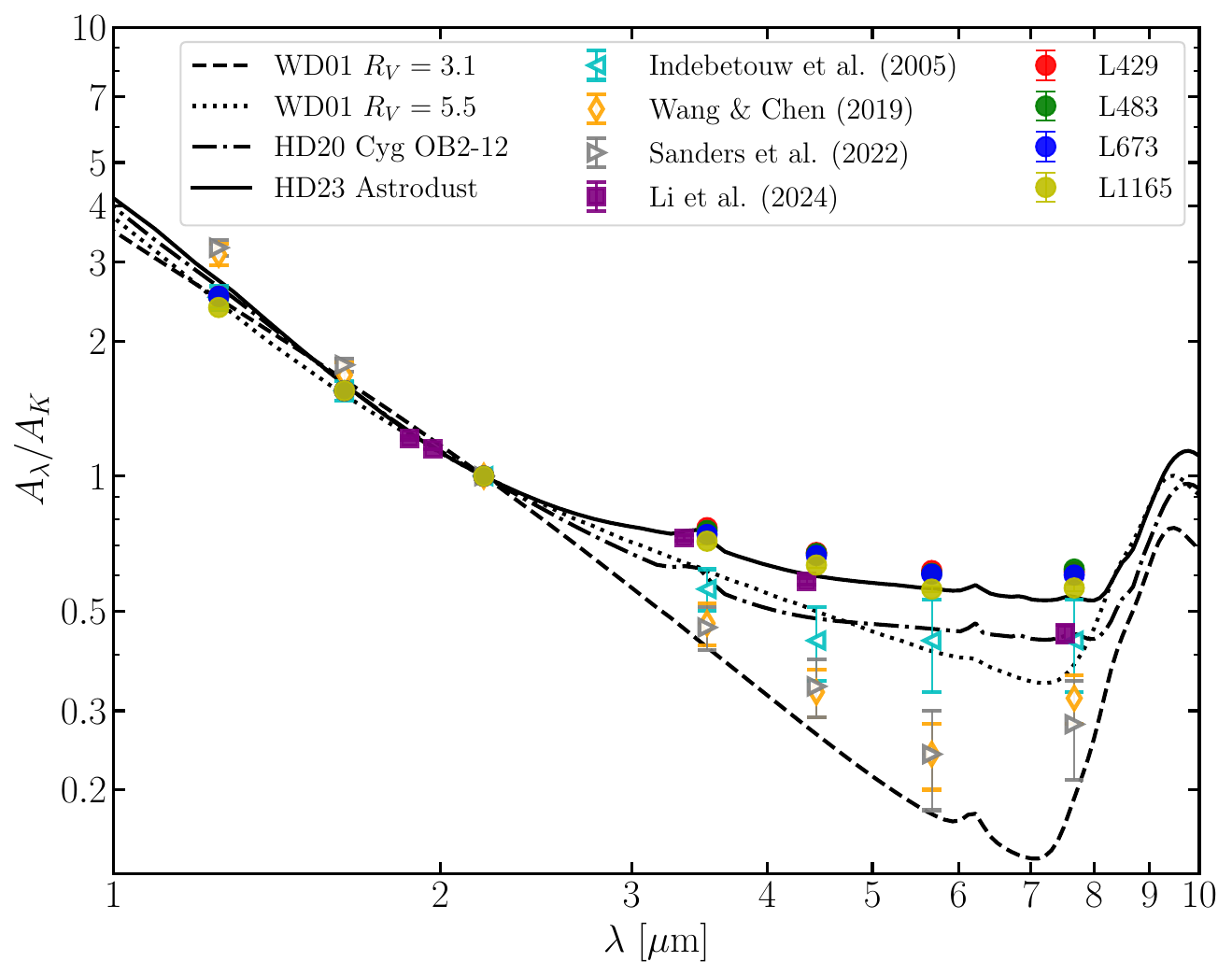}
\caption{The extinction law $A_\lambda/A_K$ for the four dense cloud cores L429, L483, L673, and L1165 (colored circles) alongside results from \citet[][WD01]{Weingartner2001} for $R_V=3.1$ (dashed line) and $R_V=5.5$ (dotted line),  \citet[][HD20 Cyg OB2-12]{Hensley2020} (dash-dotted line),  \citet[][HD23 Astrodust]{Hensley2023} (solid line). Some other observational results are shown as colored markers for comparison (\citealt{Indebetouw2005}, \citealt{Wang2019}, \citealt{Sanders2022}, \citealt{Li2024b}).
\label{fig:ext_curve}}
\end{figure*}

In this study, we find that MIR extinction law in dense molecular clouds is consistently flat, exhibiting a flatter profile than all previously observed extinction laws (see Figure \ref{fig:cer_curve} and \ref{fig:ext_curve}). A key advantage of using UKIDSS data is its ability to probe deeper extinctions, reaching depths of up to $A_V \sim 60$ mag, which significantly exceeds the capabilities of the 2MASS photometric system used in many earlier studies. Although the choice of photometric system—whether UKIDSS or 2MASS—can affect the determination of the extinction law, its impact on the color-excess ratios would be minimal. Thus our results can be directly compared with previous studies. Similar studies, such as those by \cite{Roman2007} and \cite{Ascenso2013}, also measured MIR extinction in dense cloud cores, specifically B59 and FeSt 1-457. These studies, which used deep NIR observations along with Spitzer data, found that the extinction law in these cores is appreciably flatter than the WD01 $R_V$=5.5 model. As evident from Figure \ref{fig:cer_curve}, our measured color-excess ratio curves demonstrates an even flatter profile for dark cloud cores compared to the findings of \cite{Ascenso2013}.
		
Recently, \cite{Li2024b} investigated the infrared extinction law in the dense cloud M16 using James Webb Space Telescope (JWST) data, presenting the findings in terms of the color excess ratio $E{\rm (F090W-\lambda)}/E{\rm (F090W-F200W)}$ for JWST bands. To enable direct comparison with our results, we convert their JWST color excess ratios to $A_\lambda/A_K$ using the WD01 $R_V=5.5$ model coefficients ($A_{\rm F090W}:A_{\rm F200W}:A_K=4.55:1.15:1$). This choice is supported by the good agreement between WD01 $R_V=5.5$ model and observed NIR color-excess ratios in dense environments as shown in Figure \ref{fig:cer_curve}. The resulting converted $A_\lambda/A_K$ values are displayed as purple squares in Figure \ref{fig:ext_curve}. Notably, the observed extinction law $A_\lambda/A_K$ of the four dense cores in this work are flatter than the results of \cite{Li2024b}. It should be noted that \cite{Li2024b} used $\rm (F090W-F200W)$ as the reference color which can only probe extinction depths up to $A_V\sim$ 23 mag, whereas this work use $(H-K)$ as the reference color, enabling the exploration of deeper extinctions reaching $A_V \sim 40-60$ mag. 
		
\cite{Flaherty2007} compared the IRAC-only color-excess ratio $E([3.6]-[4.5])/E([4.5]-[5.8])$ across five star-forming regions. They found that the Ophiuchus region exhibited a significantly lower ratio of 1.25$\pm$0.13, compared to values of approximately 2.5 in other regions. In our study of four dense molecular clouds, we calculated $E([3.6]-[4.5])/E([4.5]-[5.8])$ values ranging from 1.12 to 1.53, derived from the $\beta_\lambda$ values in Table \ref{tab:results}. These results are remarkably similar to those obtained by \cite{Flaherty2007} for the Ophiuchus region. This concordance further corroborates the observation that the MIR extinction law in dense regions can exhibit significant flatness, underlining the distinctive nature of extinction properties in dense environments.
		
One interesting trend noted by \cite{McClure2009} is the slight flattening of the MIR extinction law with increasing extinction. The most pronounced flattening trend was reported by \cite{Chapman2009}, who observed that the MIR extinction law becomes obviously flatter with increasing extinction in star-forming regions, although their results for low extinctions were subject to large uncertainties. Another interesting finding by \cite{Cambresy2011} was the detection of a transition in the extinction law at $A_V\sim$ 20 mag, where the extinction law became noticeably flatter for $A_V>20$ mag. However, this transition has not been consistently observed in other studies and this work. Overall, our work confirms the flatness of the MIR extinction law in dense molecular clouds, revealing an even flatter profile than all previously observed MIR extinction laws. 
		
\subsection{Implications for Grain Sizes}
		
The flat MIR extinction curve observed in the dense ISM is frequently interpreted as evidence for the presence of large dust grains \citep{Weingartner2001,Gao2013}. While dust grains in the diffuse ISM are typically sub-micron in size, they can grow significantly larger in cold dense cores through two main mechanisms: the formation of ice mantles and coagulation by grain-grain collisions \citep{Ossenkopf1994,Ormel2011,Kohler2015}. While ice mantles do form in these cold environments, their primary spectral features, such as the H$_2$O ice band at 3.0 $\mu$m fall outside the IRAC bands, making them unlikely to significantly affect the continuous extinction in our observed wavelength range. Instead, grain coagulation appears to be the dominant mechanism to more substantial changes in grain size distribution, which directly influences the observed MIR extinction law \citep{McClure2009}. This interpretation is supported by various studies: \cite{Wang2015a,Wang2015b} successfully explained the flat MIR extinction law by modifying the WD01 dust model to include micron-sized graphite or H$_2$O ice grains, while the presence of such large grains has been independently confirmed through observations of the coreshine effect  the scattering of MIR light by large dust grains \citep{Pagani2010,Steinacker2010}. The micron-sized grains produce grey extinction (i.e., extinction that does not vary significantly with wavelength) in the MIR, which is consistent with our observed flat extinction curves in the four dense clouds studied.
		
However, while the presence of micron-sized grains provides a straightforward explanation for the flat MIR extinction, other models suggest that large grain sizes may not be the only explanation. For example, the Astrodust model proposed by \cite{Hensley2023} can also reproduce the flat MIR extinction curve using typical grain sizes by carefully modeling the dielectric properties and size distribution of the grains. As shown in Figures \ref{fig:cer_curve} and \ref{fig:ext_curve}, our observed color-excess and extinction curves align more closely with the Astrodust model, suggesting that it is possible to explain the flat MIR extinction without invoking a large population of micron-sized grains. In addition, \cite{Voshchinnikov2017} demonstrated that using a combination of different dust compositions and maintaining a classical grain size distribution can also reproduce the flat extinction observed toward the Galactic center. This highlights the fact that the flat MIR extinction curve could result from a combination of grain composition, size distribution, and optical properties, not just the presence of large grains.
		
One of the limitations of the micron-sized grain hypothesis is that if grain growth were extensive enough to produce a large population of micron-sized grains, we would expect to see more dramatic changes in the extinction law at shorter wavelengths, such as in the NIR. Large grains should have a significant impact in the NIR extinction law, yet observations show that while the extinction law flattens in the MIR, the NIR extinction does not flatten as dramatically. As mentioned in Section \ref{sec:nir_ratios}, the observed NIR extinction in this work aligns closely with previously observed NIR extinction laws. This suggests that grain growth in dense clouds may be more limited than the hypothesis of extensive micron-sized grain populations would imply.
		
\begin{figure*}[ht!]
\centering
\includegraphics[scale=0.5]{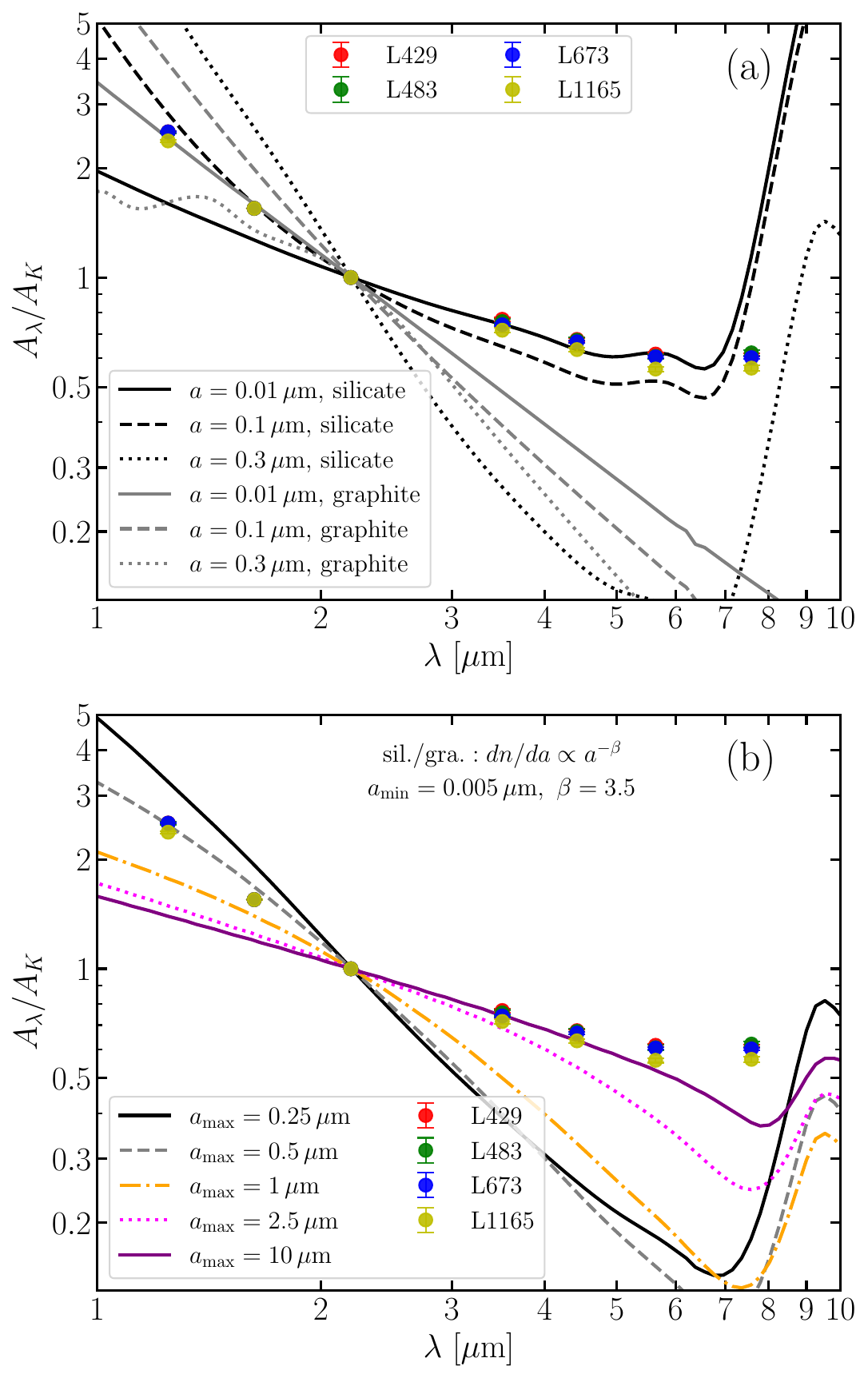}
\caption{\emph{Panel (a)} The extinction law $A_\lambda/A_K$ calculated from spherical grains with radii $a=$ 0.01\mum\ (solid lines), 0.1\mum\ (dashed lines), and 0.3\mum\ (dotted lines) of silicate (black) and graphite (grey) composition. The optical constants of silicate and graphite are taken from \cite{Draine1984}. \emph{Panel (b)} The extinction law $A_\lambda/A_K$ calculated from a mixture of silicate and graphite with a power-law size distribution $dn/da\propto a^{-3.5}$ \citep{Mathis1977}, with $a_{\rm min}=0.005\mum$ and $a_{\rm max}$ =0.25\mum\ (black solid line), 0.5\mum\ (grey dashed line), 1.0\mum\ (orange dotdashed line), 2.5\mum\ (magenta dotted line), and 10\mum\ (purple solid line). The colored circles present the observed extinction $A_\lambda/A_K$ of the four dense cores L429, L483, L673, and L1165.
\label{fig:model}}
\end{figure*}

The comparison of our observed average extinction curve from the four dense cloud cores with various model extinction laws as shown in Figure \ref{fig:model} further supports this view. When comparing the observed extinction law to the models for silicate and graphite grains of different sizes, we find that the observed curves with radius of 0.01\mum\ are flatter than the extinction predicted by models with larger grains (e.g., 0.1\mum\ and 0.3\mum). When considering a mixture of silicate and graphite grains \citep{Draine1984} with a power-law size distribution (i.e., $dn/da\propto a^{-3.5}$ in the range of $a_{\rm min}=0.005\mum$ and $a_{\rm max}$ \citep{Mathis1977}), the resulting extinction curve becomes flatter as the maximum grain radius $a_{\rm max}$ increases from 0.25\mum\ to 10\mum. However, even with $a_{\rm max}$=10\mum, the modeled extinction curve does not fully match the observed curve, especially in the IRAC [5.8] and [8.0] bands.
		
This suggests that grain growth in dense ISM is likely a complex process, with a range of grain sizes present. The presence of large grains may coexist with smaller grains, and both populations could contribute to the overall extinction properties. Micron-sized grains may play a role in flattening the MIR extinction, but they are not the sole factor. The dielectric properties of the grains, along with the size distribution and composition, are likely important contributors to the observed extinction curve in dense ISM. Therefore, while micron-sized grains offer a compelling explanation for the flatness of the MIR extinction law, the full picture is likely more nuanced. Both large grains and the intrinsic optical properties of smaller grains could work in tandem to produce the flat MIR extinction observed in our study. This highlights the need for further research into the composition, size distribution, and grain growth to better understand the extinction law in dense environments.
		
\section{Summary} \label{sec:summary}
		
In this study, we measure the infrared extinction law in four nearby isolated molecular cloud cores, L429, L483, L673, and L1165, using deep NIR data from UKIDSS and MIR data from Spitzer. We analyze the color-color diagrams and determine the color-excess ratios $E(J-H)/E(H-K)$ and $E(K-\lambda)/E(H-K)$. The NIR color-excess ratios $E(J-H)/E(H-K)$ for three of the four clouds—L429, L483, and L673—are remarkably consistent, with values around 1.75. These results align well with previous studies, such as \cite{Indebetouw2005}. However, L1165 stands out as an exception, presenting a significantly lower value of 1.5, suggesting potential effects by the presence of young stars and outflows.
		
Our findings reveal the flattest MIR extinction curve observed to date in the four dense cores, surpassing most previously reported MIR extinction laws across various ISM. The MIR extinction curves across these cores are relatively uniform, exhibiting minimal variability. Interestingly, the flatness of the MIR extinction law in these four dense cloud cores closely matches the predictions of the Astrodust model proposed by \cite{Hensley2023}. The Astrodust model, which was developed primarily for dust in the diffuse ISM, provides a reasonable fit to the observed extinction curves of cloud cores in this work. This suggests that the flat MIR extinction law can be explained not only by the presence of large dust grains but also by specific dust compositions and grain size distributions that influence the optical properties of dust in dense environments.

\section*{Acknowledgments}
We would like to thank the anonymous referee for the very helpful comments that improved this letter.  This work is supported by the National Natural Science Foundation of China (NSFC) through project Nos. 12403026, 12133002, 12173034, and 12322304, the National Key R\&D program of China 2022YFA1603102 and 2019YFA0405500. B.Q.C. acknowledges the National Natural Science Foundation of Yunnan Province 202301AV070002 and the Xingdian talent support program of Yunnan Province. X.C. thanks to Guangdong Province Universities and Colleges Pearl River Scholar Funded Scheme (2019).

\vspace{5mm}
\facilities{UKIRT, Spitzer (IRAC)}
\software{Astropy \citep{astropy2018}, LtsFit \citep{Cappellari2014}, TOPCAT \citep{Taylor2005}.}
		
\bibliography{ir_cores}{}
\bibliographystyle{aasjournal}
		
\begin{figure*}[ht!]
\centering
\includegraphics[scale=0.32]{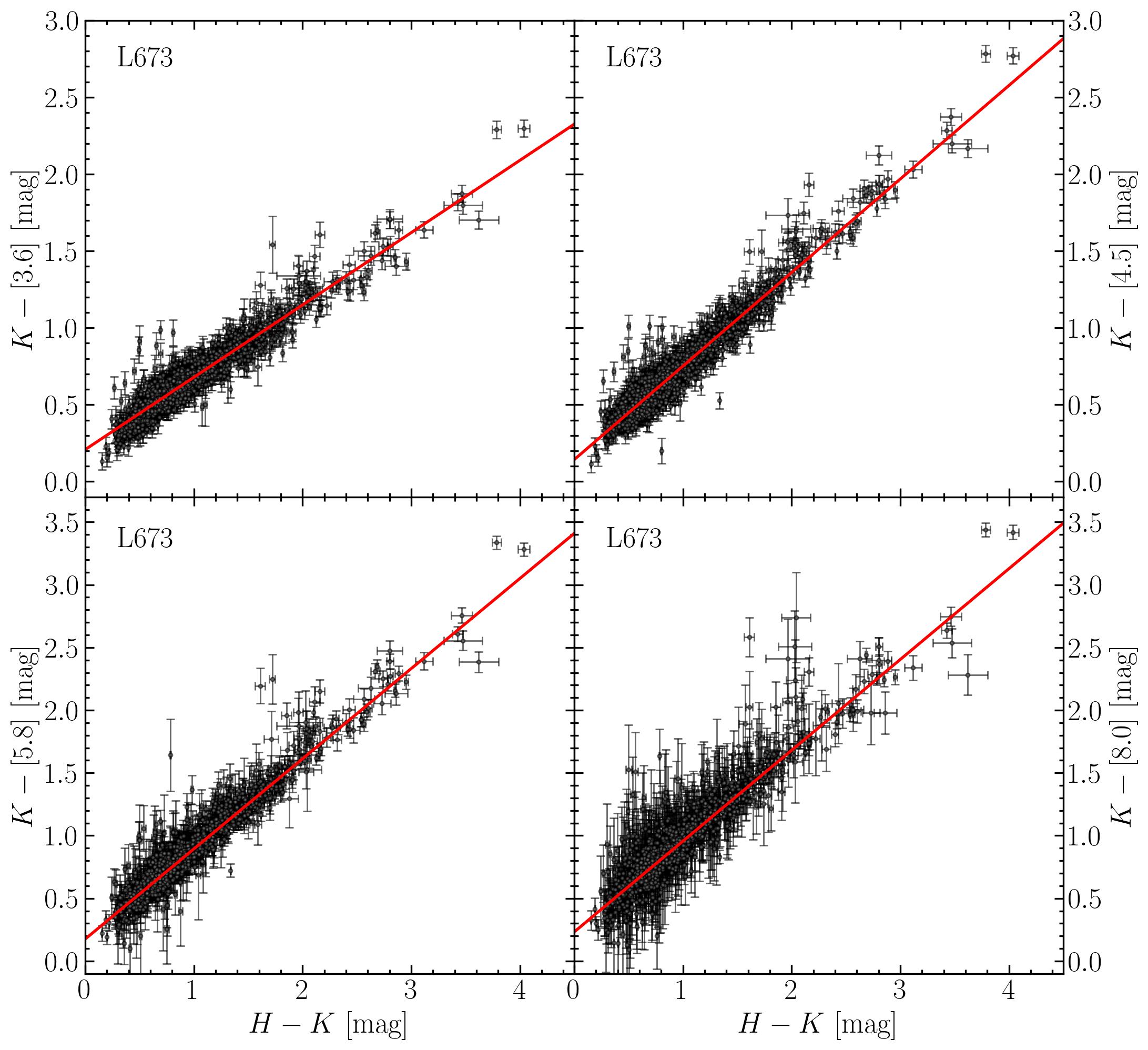}
\caption{Color-color diagrams of $(K-\lambda)$ versus $(H-K)$ for UKIDSS and Spitzer/IRAC for L673, where $\lambda$ corresponds to IRAC [3.6] (\emph{top left}), [4.5] (\emph{top right}), [5.8] (\emph{lower left}), and [8.0] (\emph{lower right}). The red solid lines represent the best-fit linear relations for each band, as determined by the \texttt{LTS\_LINEFIT} method. \label{fig:mir_ratios_L673}}
\end{figure*}

\appendix
\renewcommand*\thetable{\Alph{section}.\arabic{table}}
\renewcommand*\thefigure{\Alph{section}\arabic{figure}}
\onecolumngrid 
\setcounter{table}{0}
\setcounter{figure}{0}
\section{NIR Color Excess Ratios and Extinction Law Parameters } \label{sec:app_nir}
Table \ref{tab:nir_cer} presents the NIR extinction parameters derived for the dense molecular cores L429, L483, L673, and L1165. For each core, we determine the color excess ratio $E(J-H)/E(H-K)$ in Section \ref{sec:nir_ratios}. Assuming a power-law form for the NIR extinction, $A_\lambda\propto\lambda^{-\alpha}$, the color excess ratio can be converted to the power-law index $\alpha$ through the relationship:
\begin{equation}\label{equ:nir_cer}
\frac{E(J-H)}{E(H-K)}=\frac{\left(\frac{\lambda_H}{\lambda_J}\right)^\alpha-1}{1-\left(\frac{\lambda_H}{\lambda_K}\right)^\alpha},
\end{equation}
where $\lambda_J$, $\lambda_H$, and $\lambda_K$ represent the effective wavelengths of the UKIDSS $J$, $H$, and $K$ bands, respectively, as specified in Table \ref{tab:results}. Table \ref{tab:nir_cer} presents the derived power-law indices $\alpha$ and the calculated NIR extinction ratios $A_J/A_K$ and $A_H/A_K$. For comparative analysis, we include corresponding values from \cite{Indebetouw2005}, the WD01 models with $R_V=3.1$ and $R_V=5.5$ \citep{Weingartner2001}, and the Astrodust model developed by \cite{Hensley2023}. 
		
\begin{table*}
\centering
\caption{Near-infrared extinction law parameters derived for the four dense molecular cores compared with previously published values from the literature.}
\label{tab:nir_cer}
\begin{tabular}{ c  c  c  c  c }
\hline
& $E(J-H)/E(H-K)$ & $\alpha$ & $A_J/A_K$ & $A_H/A_K$ \\ \hline
L429             & 1.758$\pm$0.010   & 2.021$\pm$0.021 & 2.965$\pm$0.034 & 1.712$\pm$0.010 \\
L483             & 1.749$\pm$0.022   & 2.002$\pm$0.047 & 2.935$\pm$0.074 & 1.704$\pm$0.021 \\
L673             & 1.764$\pm$0.008   & 2.034$\pm$0.017 & 2.985$\pm$0.027 & 1.718$\pm$0.008 \\
L1165            & 1.499$\pm$0.018   & 1.430$\pm$0.045 & 2.157$\pm$0.052 & 1.463$\pm$0.017 \\ 
\hline
\cite{Indebetouw2005} & 1.73          & --            & 2.50$\pm$0.15 & 1.55$\pm$0.08 \\ 
WD01 $R_V=3.1$  & 1.33          & 1.58          & 2.49          & 1.64          \\ 
WD01 $R_V=5.5$   & 1.76          & 1.77          & 2.44          & 1.52          \\
HD23 Astrodust        & 1.89          & 1.93          & 2.74          & 1.60          \\ \hline
\end{tabular}
\end{table*}

\section{Color-color diagrams}\label{sec:app_mir_cer}
\begin{figure*}[ht!]
\centering
\includegraphics[scale=0.32]{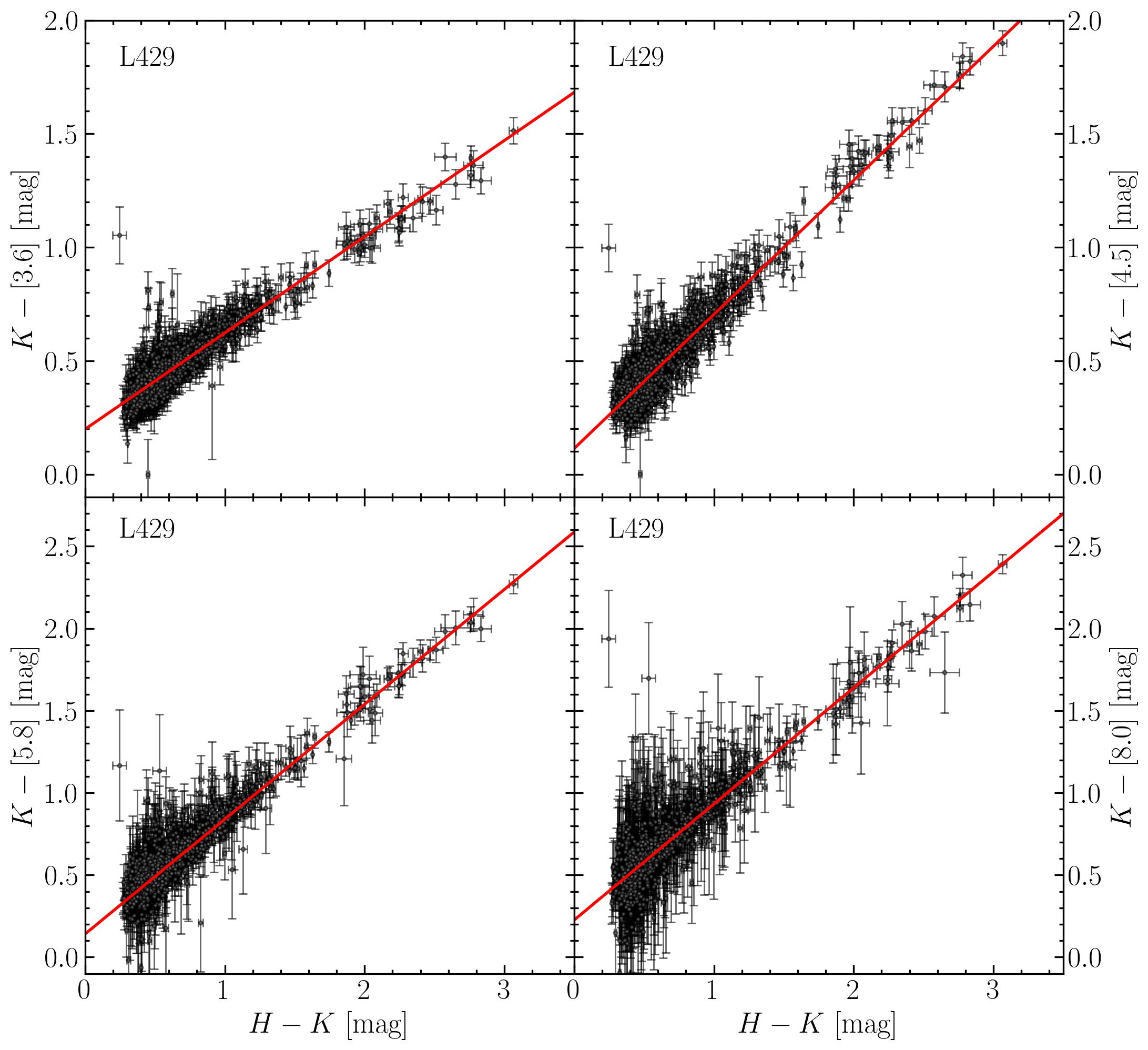}
\caption{Analogous to Figure \ref{fig:mir_ratios_L673}, but for the L429 core.
\label{fig:mir_ratios_L429}}
\end{figure*}

\begin{figure*}[ht!]
\centering
\includegraphics[scale=0.32]{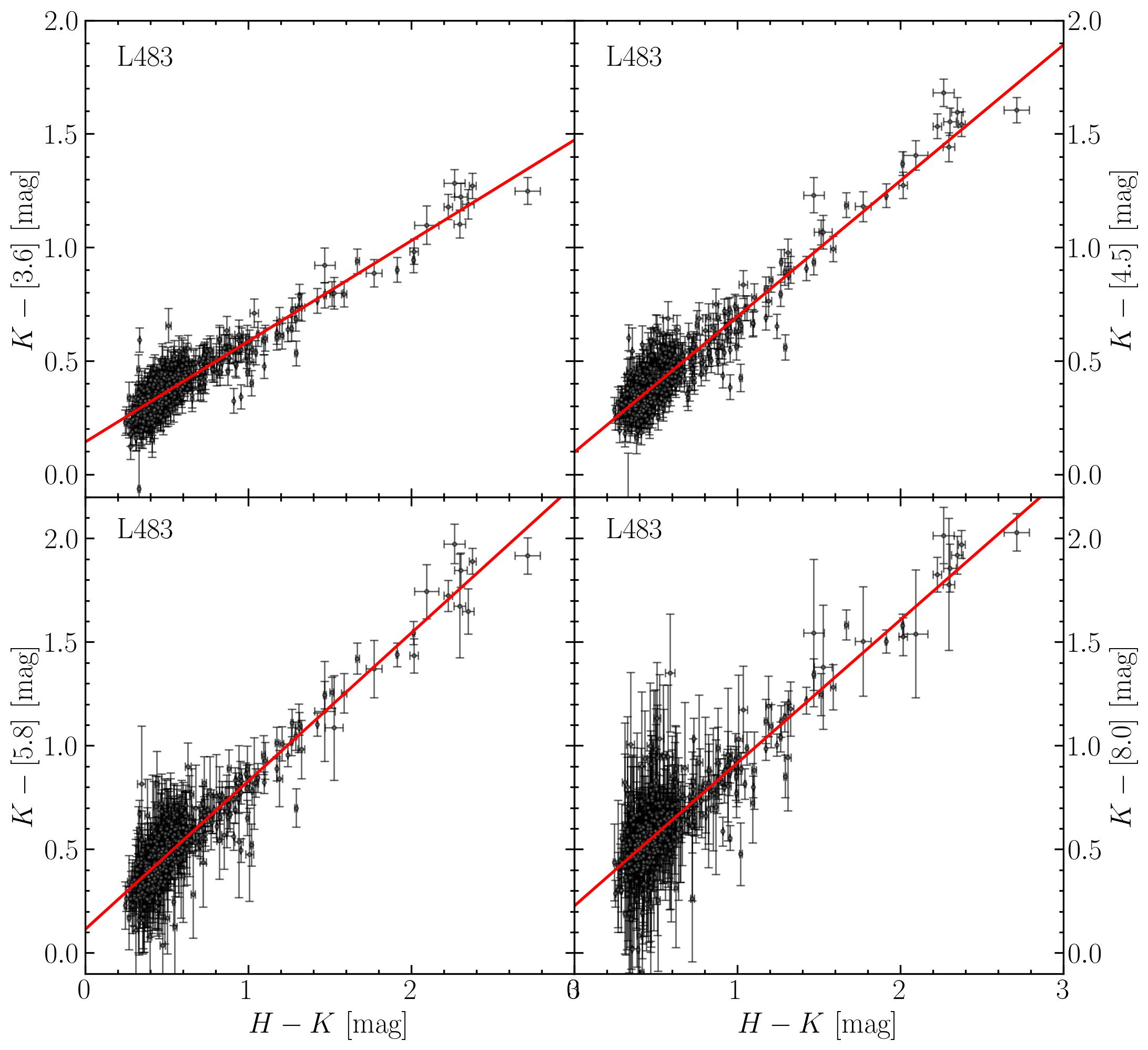}
\caption{Analogous to Figure \ref{fig:mir_ratios_L429}, but for the L483 core.
\label{fig:mir_ratios_L483}}
\end{figure*}

\begin{figure*}[ht!]
\centering
\includegraphics[scale=0.32]{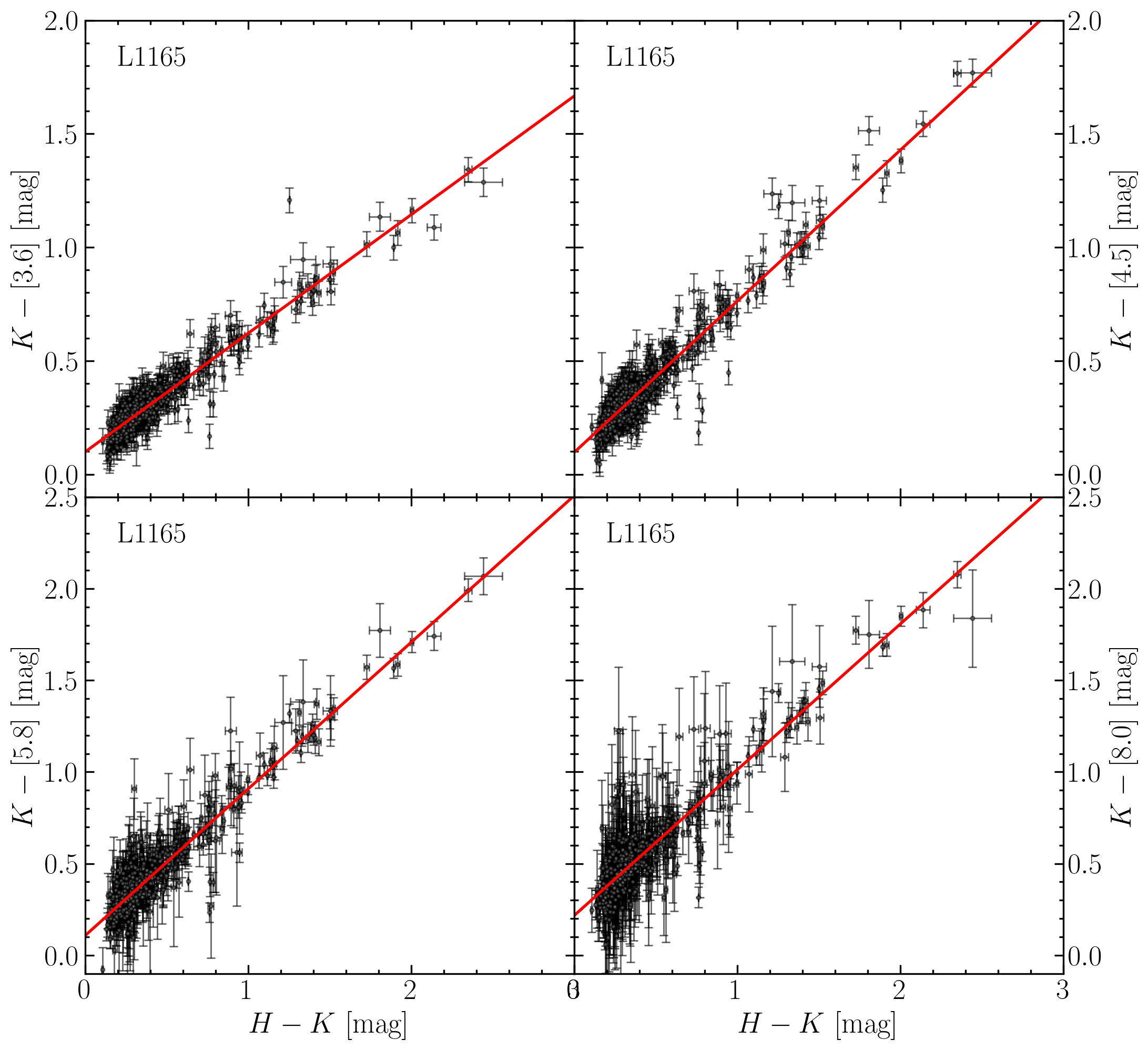}
\caption{Analogous to Figure \ref{fig:mir_ratios_L429}, but for the L1165 core.
\label{fig:mir_ratios_L1165}}
\end{figure*}

Figures \ref{fig:mir_ratios_L429}-\ref{fig:mir_ratios_L1165} show the color-color diagrams $(K-\lambda)$ versus $(H-K)$ for UKIDSS and Spitzer/IRAC for the dense molecular cloud cores L429, L483, and L1165. The captions for these figures are analogous to Figure \ref{fig:mir_ratios_L673} outlined in Section \ref{sec:EHK}.  The $K-[8.0]$ relationships exhibit larger scatter compared to the other three bands, attributable to two primary factors: increased photometric uncertainties in the [8.0] band due to lower sensitivity and increased background emission, as well as possible contamination from polycyclic aromatic hydrocarbon (PAH) emission features.
		
\clearpage
\end{CJK*}
\end{document}